\documentclass[aps,10pt,prb,twocolumn,superscriptaddress]{revtex4-2}
\usepackage{graphicx}
\usepackage{amsmath}
\usepackage{amsfonts,hyperref}
\usepackage{amssymb,mathrsfs}
\usepackage{upgreek}
\usepackage{bbold,yfonts}

\newcommand{\bs}[1]{\boldsymbol{#1}}

\begin{document}

\title{Anti-Poiseuille flow in neutral graphene}

\author{B.N. Narozhny}
\affiliation{\mbox{Institut for Theoretical Condensed Matter Physics, Karlsruhe Institute of 
Technology, 76128 Karlsruhe, Germany}}
\affiliation{National Research Nuclear University MEPhI (Moscow Engineering Physics Institute),
  115409 Moscow, Russia}
\author{I.V. Gornyi}
\affiliation{\mbox{Institut for Quantum Materials and Technologies, Karlsruhe Institute of
Technology, 76021 Karlsruhe, Germany}}
\affiliation{\mbox{Institut for Theoretical Condensed Matter Physics, Karlsruhe Institute of
Technology, 76128 Karlsruhe, Germany}}
\affiliation{Ioffe Institute, 194021 St. Petersburg, Russia}
\author{M. Titov}
\affiliation{Radboud University Nijmegen, Institute for Molecules and Materials, NL-6525 AJ
Nijmegen, The Netherlands}

\date{\today}

\begin{abstract}
  Hydrodynamic flow of charge carriers in graphene is an energy flow
  unlike the usual mass flow in conventional fluids. In neutral
  graphene, the energy flow is decoupled from the electric current,
  making it difficult to observe the hydrodynamic effects and measure
  the viscosity of the electronic fluid by means of electric current
  measurements. In particular, we show that the hallmark Poiseuille
  flow in a narrow channel cannot be driven by the electric field
  irrespective of boundary conditions at the channel edges.
  Nevertheless one can observe nonuniform current densities similarly
  to the case of the well-known ballistic-diffusive crossover. The
  standard diffusive behavior with the uniform current density across
  the channel is achieved under the assumptions of specular scattering
  on the channel boundaries. This flow can also be made nonuniform by
  applying weak magnetic fields. In this case, the curvature of the
  current density profile is determined by the quasiparticle
  recombination processes dominated by the disorder-assisted
  electron-phonon scattering -- the so-called supercollisions.
\end{abstract}

\maketitle

Electronic hydrodynamics has attracted substantial experimental and
theoretical attention in recent years \cite{pg,rev,luc}. Hydrodynamic
flows in two-dimensional (2D) materials can now be observed directly
using several imaging techniques
\cite{fink,halb,ihn,sulp,zel19,uri,imm,imh,vool,young,zel}. Two of
these experiments \cite{imm,imh} were focusing on the Poiseuille flow,
the simplest manifestation of viscous hydrodynamics in conventional
fluids \cite{dau6}.

The Poiseuille flow \cite{poi,poise,dau6} is a pressure-induced flow
in a pipe or between parallel plates. The latter is equivalent to a 2D
flow in a narrow channel (with the length $L$ much greater than the
width $W$). In the middle of the channel (away from both of its ends)
the flow velocity is directed along the channel and depends only on
the transverse coordinate. In that case, the hydrodynamic equations
admit a simple solution with the parabolic velocity profile and the
flow rate (discharge) that is proportional to the third power of
the channel width (for a three-dimensional flow through a pipe -- the
fourth power of the radius, which is especially important in
hematology \cite{hema}).

The possibility for an electronic system to exhibit the Poiseuille
flow in a narrow wire was first pointed out by Gurzhi
\cite{gurzhi1,gurzhi2,gurzhi}. Recently, similar behavior has been a
subject of intense theoretical
\cite{mol95,pol15,fl0,ale,moo,pol17,mr2,cfl,ady2,mey20,glazman20,sven1}
and experimental
\cite{mol95,imm,imh,vool,mac,geim5,geim2,young,gus18,gus20,gus21,rai20,gupta,var20,goo2,jao}
research in the context of electronic transport in high-mobility 2D
materials. In contrast to conventional fluids, the electronic flow is
affected not only by viscous effects, but also by weak disorder
scattering and is characterized by a typical length scale known as the
Gurzhi length \cite{moo,pol17,mr2,cfl,sven1}
\begin{equation}
\label{gl}
\ell_G = \sqrt{\nu\tau_{\rm dis}}.
\end{equation}
Here $\nu$ is the kinematic viscosity \cite{dau6,luc,geim1,geim4,me2}
and $\tau_{\rm dis}$ is the disorder mean free time. The resulting
current profile is given by the catenary curve approaching the
parabola in the limit ${\ell_G\gg{W}}$.

Nonuniform hydrodynamic flow in a narrow channel has to be contrasted
with a conventional ballistic flow that in the case of realistic
boundary conditions \cite{imm,bee} can also be nonuniform. Assuming
rough edges, where electrons scatter off in all directions with equal
probability (``diffusive scattering''), bulk impurity scattering
competes with boundary effects leading to a ballistic-diffusive
crossover. If the mean free path is much smaller than the channel
width, $\ell_{\rm dis}\ll W$, then the electric current density is
uniform, except for the small regions close to the edges. Reducing the
channel width leads to the appearance of a curved current profile that
is visually similar to the Poiseuille flow (with the maximum curvature
corresponding to both length scales being of the same order of
magnitude). In doped graphene this was observed in the recent imaging
experiment \cite{imm}.

Physics of neutral graphene \cite{imh,kim1,gal} is more
intricate. Here the electronic system is nondegenerate and both
graphene bands contribute to transport on equal footing. Due to
linearity of the Dirac spectrum, the Auger processes are kinematically
suppressed and to the leading approximation the number of particles in
each band is conserved independently \cite{rev,luc,alf,bri}. Another
consequence of the peculiar kinematics of Dirac fermions in graphene
is the so-called ``collinear scattering singularity''
\cite{msf,mfss,kash,mus,bri,schutt,drag2,hydro0} that gives rise to
the ``three-mode approximation'' allowing one to solve the kinetic
equation and derive the hydrodynamic theory \cite{hydro0,hydro1,me1}.
The key feature of the resulting description is that the hydrodynamic
flow in graphene is the {\it flow of energy} rather than mass in
conventional fluids or charge in Ohmic conductors
\cite{rev,luc,hydro1,me1}. Precisely at charge neutrality and in the
absence of external magnetic field, the hydrodynamic energy flow is
completely decoupled from the electric current. In an infinite system
the latter exhibits usual Ohmic behavior with the dominant
contribution to the mean free path coming from electron-electron
interaction \cite{gal,kash,mfss,hydro1,fog,me1,me3}. It is then
reasonable to expect that in a narrow channel this current should
exhibit the above ballistic-diffusive crossover with the only
difference being the microscopic nature of the mean free path.

Hydrodynamic flows in neutral graphene were recently studied
experimentally with the help of nanoscale magnetic imaging
\cite{imh}. The authors reported measurements of inhomogeneous
electric current density interpreting them in terms of the
Poiseuille flow. Assuming that the curvature of the current density
profile was determined by viscosity, the authors proceeded to extract
the shear viscosity in graphene at and close to charge neutrality. The
resulting values appeared to be in a surprisingly good agreement with
the theoretical calculations of Ref.~\cite{me2}.

What exactly is the Poiseuille flow and can it be used as a hallmark
of hydrodynamic behavior? The Poiseuille flow is a particular solution
to the Navier-Stokes equation \cite{dau6} in the case where a viscous,
incompressible fluid is constrained by (straight and infinitely long)
stationary boundaries. The problem is usually solved under the
assumption of the so-called {\it no-slip boundary conditions}, i.e.
the vanishing flow velocity at the boundaries. Then the Navier-Stokes
equation becomes an ordinary second-order differential equation
yielding the standard parabolic velocity profile. The solution can be
extended to the case of more general Maxwell's boundary conditions
\cite{max} with a finite slip length \cite{ks19}. The limit of the
infinite slip length however (i.e., with no-stress boundary
conditions) does not admit any solutions for the Poiseuille
problem. In other words: {\it a pressure-induced viscous flow in a
  pipe cannot be homogeneous}. On the contrary, an inviscid fluid is
described by the Euler equation \cite{dau6}, which is a nonlinear,
first-order differential equation. As such, it does not require
boundary conditions on the longitudinal (along the boundary) component
of the velocity and allows for homogeneous solutions. Hence, the
Poiseuille flow can be used as a hallmark of viscosity.

Adapting the above arguments to electronic transport is
straightforward for single-band, Fermi-liquid-like systems, such as
doped graphene. Here all physical quantities are determined by the
Fermi energy, all macroscopic currents are physically equivalent
and can be represented by a single vector quantity, the velocity
$\bs{u}$. In the ``hydrodynamic regime'', i.e., if the
electron-electron interaction is the dominant scattering mechanism in
the problem, $\ell_{ee}\ll\ell_{\rm dis},\ell_{e-ph},W$ (in the
self-evident notation), $\bs{u}$ obeys a Navier-Stokes-like equation
\cite{luc,rev,me1} and may exhibit a Poiseuille-like behavior in
a channel \cite{imm}.

{\it Electronic hydrodynamics in graphene} -- 
In a two-band system the situation is more involved. An
out-of-equilibrium (current-carrying) state may be characterized
either by the chemical potentials $\mu_\pm$ of each band, or by their
linear combinations \cite{alf,me1}
\begin{subequations}
\label{muns}
\begin{equation}
\label{mus}
\mu = (\mu_+\!+\!\mu_-)/2,
\qquad
\mu_I = (\mu_+\!-\!\mu_-)/2,
\end{equation}
conjugate to the charge and imbalance densities
\begin{equation}
\label{ns}
n = n_+ - n_-, \qquad
n_I = n_+ + n_-.
\end{equation}
\end{subequations}
In equilibrium ${\mu_I=0}$. Although macroscopic currents are no
longer equivalent \cite{hydro0,hydro1,alf,me1}, one can still
introduce the hydrodynamic velocity associating it with one (nearly)
conserved current, namely the momentum flux. In the case of linear
spectrum, the momentum flux is equivalent to the energy current. As a
result, the electric ($\bs{j}$), quasiparticle (or ``imbalance'',
$\bs{j}_I$), and energy ($\bs{j}_E$) currents in graphene can be
defined as \cite{rev,luc,me1}
\begin{equation}
\label{jlr}
\bs{j} = n\bs{u} \!+\! \delta\bs{j},
\quad
\bs{j}_I = n_I\bs{u} \!+\! \delta\bs{j}_I,
\quad
\bs{j}_E = {\cal W}\bs{u},
\end{equation}
where ${\cal W}$ is the enthalpy density and $\delta\bs{j}$ and
$\delta\bs{j}_I$ are the dissipative corrections, see Eqs.~(\ref{djs})
below and the Appendix. In the degenerate limit ${\mu\gg{T}}$ the
dissipative corrections vanish \cite{me1,me3} justifying the
applicability of the above single-band picture to doped graphene. At
charge neutrality ${n=0}$, the electric and energy currents in
Eq.~(\ref{jlr}) appear to be decoupled \cite{me1}.

The quasiparticle currents $\bs{j}$ and $\bs{j}_I$ satisfy the
continuity equations \cite{rev,luc,me1,meig1}
\begin{subequations}
\label{heqs}
\begin{equation}
\label{cen1}
\partial_t n + \bs{\nabla}\!\cdot\!\bs{j} = 0,
\end{equation}
\begin{equation}
\label{ceni1}
\partial_t n_I + \bs{\nabla}\!\cdot\!\bs{j}_I = - \frac{n_I\!-\!n_{I,0}}{\tau_R}
=-\frac{12\ln2}{\pi^2}\frac{n_{I,0}\mu_I}{T\tau_R},
\end{equation}
where ${n_{I,0}=\pi{T}^2/(3v_g^2)}$ is the equilibrium value of the
total quasiparticle density (i.e., at $\mu_I=0$) and $\tau_R$ is the
recombination time \cite{meg,meig1}. The hydrodynamic velocity
$\bs{u}$ satisfies the generalized Navier-Stokes equation \cite{me1}
\begin{eqnarray}
\label{eq1g}
&&
\!\!\!\!\!\!
{\cal W}(\partial_t+\bs{u}\!\cdot\!\bs{\nabla})\bs{u}
+
v_g^2 \bs{\nabla} P
+
\bs{u} \partial_t P 
+
e(\bs{E}\!\cdot\!\bs{j})\bs{u} 
=
\\
&&
\nonumber\\
&&
=
v_g^2 
\left[
\eta \Delta\bs{u}
+
en\bs{E}
+
\frac{e}{c} \bs{j}\!\times\!\bs{B}
\right]
-
{\cal W}\bs{u}/\tau_{{\rm dis}},
\nonumber
\end{eqnarray}
where $P$ and $\eta$ are the thermodynamic pressure and shear
viscosity. The full hydrodynamic equations \cite{alf,lev19} also
includes the thermal transport equation \cite{meig1}
\begin{eqnarray}
\label{eqen}
&&
T\left[\frac{\partial s}{\partial t}
+
\bs{\nabla}\!\cdot\!
\left(s\bs{u}-\delta\bs{j}\frac{\mu}{T}-\delta\bs{j}_I\frac{\mu_I}{T}\right)\right]
=
\\
&&
\nonumber\\
&&
\qquad\qquad
=
\delta\bs{j}\!\cdot\!
\left[e\bs{E}\!+\!\frac{e}{c}\bs{u}\!\times\!\bs{B}\!-\!T\bs{\nabla}\frac{\mu}{T}\right]
-
T\delta\bs{j}_I\!\cdot\!\bs{\nabla}\frac{\mu_I}{T}
\nonumber\\
&&
\nonumber\\
&&
\qquad\qquad\qquad
+
\frac{\eta}{2}\left(\nabla_\alpha u_\beta \!+\! \nabla_\beta u_\alpha
\!-\! \delta_{\alpha\beta} \bs{\nabla}\!\cdot\!\bs{u}\right)^2
\nonumber\\
&&
\nonumber\\
&&
\qquad\qquad\qquad
-
\frac{n_E\!-\!n_{E,0}}{\tau_{RE}}
+
\mu_I \frac{n_I\!-\!n_{I,0}}{\tau_R}
+
\frac{{\cal W}\bs{u}^2}{v_g^2\tau_{\rm dis}},
\nonumber
\end{eqnarray}
\end{subequations}
which is typically used in hydrodynamics \cite{dau6} instead of the
continuity equation representing energy conservation. Here $n_{E,0}$
denotes the equilibrium value of the energy density similarly to
$n_{I,0}$ (i.e., at ${\mu_I=0}$) and $\tau_{RE}$ is the energy
relaxation time (due to, e.g., supercollisions \cite{meig1}). The last
three terms in Eq.~(\ref{eqen}) represent energy relaxation, entropy
increase due to quasiparticle recombination, and local heating due to
impurity scattering.

Consider now linear response transport in the channel geometry
(see Refs.~\cite{imm,imh} for experimental realization) at charge
neutrality (${n=0}$) in the steady state. Linearizing the hydrodynamic
equations (\ref{heqs}), we obtain \cite{meig1}

\begin{subequations}
\label{hydrolin1}
\begin{equation}
\label{cen2}
\bs{\nabla}\!\cdot\!\delta\bs{j} = 0,
\end{equation}
\begin{equation}
\label{ceni2}
n_{I,0}\bs{\nabla}\!\cdot\!\bs{u} + \bs{\nabla}\!\cdot\!\delta\bs{j}_I 
= -(12\ln2/\pi^2)n_{I,0}\mu_I/(T\tau_R),
\end{equation}
\begin{eqnarray}
\label{nseqlin1}
\bs{\nabla} \delta P
=
\eta \Delta\bs{u}
+
(e/c)\delta\bs{j}\!\times\!\bs{B}
-
3P\bs{u}/(v_g^2\tau_{{\rm dis}}),
\end{eqnarray}
\begin{equation}
\label{tteqlin1}
3P\bs{\nabla}\!\cdot\!\bs{u} = - 2\delta{P}/\tau_{RE},
\end{equation}
\end{subequations}
where we have used the ``equation of state'' \cite{me1}
\[
{\cal W}=3{P}=3{n}_E/2.
\]
Here we follow the standard approach \cite{dau6} where the
thermodynamic quantities are replaced by the corresponding equilibrium
functions of the hydrodynamic variables. Equations (\ref{hydrolin1})
should be solved for the unknowns $\bs{u}$, $\mu_I$, and $\delta P$
keeping the rest of the quantities, e.g., $n_{I,0}$, $P$, and
$T$, constant (the dissipative corrections $\delta\bs{j}$,
$\delta\bs{j}_I$ are specified below).

At charge neutrality, the electric field vanishes from the linearized
Navier-Stokes equation (\ref{nseqlin1}) and hence cannot drive a
hydrodynamic flow.

{\it Channel geometry: absence of the Poiseuille flow in neutral graphene} --
The channel geometry can be modeled by an
``infinite'' strip (i.e., with the length of the sample much greater
than its width). Transport measurements are assumed to be performed in
the two-terminal scheme \cite{imm,imh} with the leads placed at the
far away ends of the channel. In the middle of the sample, the
electric current is flowing along the channel and all physical
quantities are independent of the longitudinal coordinate $x$ (this is
not true in small regions close to the leads at the ends of the
channel). At ${n=0}$, the electric current is given by the
dissipative correction ($y$ is the transverse coordinate)
\begin{subequations}
\label{strvars}
\begin{equation}
\label{jstr}
\bs{j}=\delta\bs{j} = \delta j_x(y) \bs{e}_x,
\end{equation}
automatically satisfying the continuity equation (\ref{cen2}). The
pressure is also a function of $y$
\begin{equation}
\label{pstr}
\delta P=\delta P(y)
\quad\Rightarrow\quad
\bs{\nabla}\delta P = \frac{\partial\delta P}{\partial y}\bs{e}_y,
\end{equation}
and similarly 
\begin{equation}
\label{muistr}
\mu_I=\mu_I(y)
\quad\Rightarrow\quad
\bs{\nabla}\mu_I = \frac{\partial\mu_I}{\partial y}\bs{e}_y.
\end{equation}
Projecting the Navier-Stokes equation (\ref{nseqlin1}) onto the
longitudinal direction, we find
\begin{equation}
\label{nseqlin1x}
\eta \frac{\partial^2 u_x}{\partial y^2} = \frac{3Pu_x}{v_g^2\tau_{\rm dis}}
\quad\Rightarrow\quad
u_x=0.
\end{equation}
This is a homogeneous equation that yields the trivial solution
$u_x=0$ for either the no-slip or no-stress boundary conditions. As a
result,
\begin{equation}
\label{ustr}
\bs{u} = u_y(y) \bs{e}_y
\quad\Rightarrow\quad
\bs{\nabla}\!\cdot\!\bs{u} = \frac{\partial u_y}{\partial y}.
\end{equation}
\end{subequations}

Equations (\ref{strvars}) represent the key difference between the
usual hydrodynamic flow and electronic transport in neutral
graphene. The standard Poiseuille flow is driven by the pressure
gradient. In contrast, charge carriers in graphene may be driven by
the electric field. At charge neutrality, the field term vanishes from
the Navier-Stokes equation leading to the homogeneous equation
(\ref{nseqlin1x}) for the longitudinal component of the velocity. In
other words, {\it in neutral graphene the Poiseuille flow cannot be
  driven by the electric field}. Instead, one should apply a
temperature gradient along the channel [in this case, the pressure
  gradient in Eq.~(\ref{pstr}) will acquire an $x$-component
  contributing a driving term to Eq.~(\ref{nseqlin1x})], see also
Ref.~\cite{julia}. We emphasize that this result does not depend on
microscopic details of carrier scattering off the channel edges.

What does this mean for the electric current? To clarify this
question, we have to specify the dissipative corrections
$\delta\bs{j}$ and $\delta\bs{j}_I$. Their general form was derived in
bulk graphene in Refs.~\cite{me1,me3}, see also Appendix. This
derivation relied on the specific form of the nonequilibrium
correction to the distribution function [see Eq.~(\ref{hs}) in
the Appendix] representing a natural generalization of the usual
solution to the kinetic equation in metals \cite{ziman} to the
two-band Dirac system in graphene. In a narrow channel, solutions to
the kinetic equation should be subjected to boundary conditions
\cite{bee} reflecting the nature of the electron scattering off the
channel edges. Specifically at charge neutrality, the typical
wavelength of Dirac quasiparticles is determined by temperature and
thus is much larger than the length scale of the edge roughness that
may lead to diffusive boundary scattering \cite{bee}. As a result,
specular boundary conditions can be expected to adequately describe
neutral graphene samples.

In the limit of specular scattering, the distribution
function Eq.~(\ref{hs}) satisfies the boundary conditions and the form
of the dissipative corrections remains the same as in the bulk
system. At charge neutrality, the corrections are given by
\begin{subequations}
\label{djs}
\begin{equation}
\label{lj}
\delta\bs{j} = 
\frac{1}
{e^2\tilde{R}}
\left[ e\bs{E}
+
\omega_B\bs{e}_B\!\times\!
\left(
\frac{\alpha_1\delta_I\bs{\nabla}\mu_I}{\tau_{\rm dis}^{-1}\!+\!\delta_I^{-1}\tau_{22}^{-1}}
-
\frac{2T\ln2}{v_g^2}\bs{u}\right)\!\right]\!,
\end{equation}
\begin{eqnarray}
\label{lji}
&&
\delta\bs{j}_I = 
-\frac{\delta_I}{\tau_{\rm dis}^{-1}\!+\!\delta_I^{-1}\tau_{22}^{-1}}
\frac{1}{e^2\tilde{R}}\times
\\
&&
\nonumber\\
&&
\quad
\times\!
\left[ \alpha_1\omega_B\bs{e}_B\!\times\!\bs{E}
\!+\!
\frac{2T\ln2}{\pi}e^2R_0\bs{\nabla}\mu_I
\!+\!
\alpha_1\omega_B^2\frac{2T\ln2}{v_g^2}
\bs{u} \right]\!,
\nonumber
\end{eqnarray}
\begin{equation}
\tilde{R}=R_0\!+\!\alpha_1^2\delta_I\tilde{R}_B.
\end{equation}
\end{subequations}
Here $R_0$ [see Eq.~(\ref{r0})] is the zero-field bulk resistivity in
neutral graphene \cite{hydro0,me1,mus},
${\tilde{R}_B\propto\omega_B^2\tau_{\rm dis}}$ is defined in
Eq.~(\ref{trb}), ${\omega_B=eBv_g^2/(2cT\ln2)}$ is the generalized
cyclotron frequency (at ${\mu=0}$), ${\alpha_1\approx2.08}$ and
${\delta_I\approx0.28}$ are detailed in Appendix ($v_g$ is the band
velocity in graphene, $c$ is the speed of light, and $e$ is the
electron charge). The parameter $\tau_{22}$ describes the integrated
collision integral, see Eqs.~(\ref{is}). Both $\tau_{\rm dis}$ and
$\tau_{22}$ are functions of the chemical potential and temperature
\cite{me1,me3,drag}.

At ${\bs{B}=0}$, the corrections (\ref{djs}) simplify. The electric
current (${e\delta\bs{j}=\bs{E}/R_0}$) is governed by Ohmic
dissipative processes and is independent of the hydrodynamic velocity.
Thus, we immediately arrive at the conclusion that in the absence of
magnetic field the resulting current density in neutral graphene with
specular boundaries is uniform \cite{me1,me3} (in contrast to
conventional hydrodynamics that does not allow for a stationary
pressure-induced flow in a channel without boundary friction
\cite{dau6}).

{\it Nonuniform flows in magnetic field} --
Now we show that even in the case of specular scattering on the
channel boundaries the electric current density can be made nonuniform
by applying weak external magnetic field. In the presence of the field
all three macroscopic currents are entangled \cite{hydro0} and one may
expect a nontrivial solution.  The electric current is still flowing
along the channel, but is accompanied by the lateral flow of
quasiparticles \cite{meg,mr1}.  Since the latter cannot leave the
sample, this flow has to vanish at both edges and (nontrivial)
homogeneous solutions are no longer allowed. In the two-fluid model of
compensated semimetals \cite{mr1,mr2,mr3,sven2} the nontrivial
inhomogeneous solution becomes possible due to quasiparticle
recombination.

Quasiparticle recombination refers to any scattering process that
violates the ``approximate'' conservation of the number of particles
in each individual band including the kinematically suppressed Auger
processes, three-particle collisions, scattering by optical phonons
\cite{fos16,lev19}, and the disorder-assisted electron-phonon coupling
(or ``supercollisions'') \cite{srl,ralph13,betz,tik18,kong,meig1}. The
resulting quasiparticle recombination is manifested by an additional
term in the continuity equation (\ref{ceni1}) for the total
quasiparticle (``imbalance'') density, first established in
Ref.~\cite{alf} in the context of thermoelectric phenomena. Recently,
recombination effects were shown to lead to linear magnetoresistance
in compensated semimetals \cite{mrexp,mr1,mr3,mr2}, giant
magnetodrag \cite{drag12,meg}, and giant nonlocality \cite{nlr,sven2}.

Supercollisions involve electron-phonon scattering in a close
proximity to an impurity. This is a second-order process where an
electron in the upper graphene band may scatter into an empty state in
the lower band while emitting a phonon and losing its momentum to the
impurity. In the reverse process, the phonon can be absorbed by an
electron in the lower band scattering into the upper band (while the
impurity compensates the momentum mismatch). Unlike the Auger or
three-particle processes, supercollisions also lead to energy
relaxation \cite{meig1}. Taking into account recombination without
energy relaxation leads to a problem: the continuity equations for
energy and imbalance densities allow only homogeneous solutions, which
are incompatible with the boundary conditions at the channel
edges. Here we show that energy relaxation due to supercollisions
provides the missing piece of the puzzle allowing one to solve the
hydrodynamic equations in graphene at charge neutrality. The solution
exhibits the inhomogeneous electric current profile in neutral
graphene samples with specular reflective boundaries subjected to weak
magnetic field. We find that the curvature of the current profile is
determined by supercollisions (by means of energy relaxation and
quasiparticle recombination) rather than viscosity. A
case of rough edges and the corresponding ballistic-diffusive
crossover will be discussed elsewhere.

Substituting Eqs.~(\ref{strvars}) into Eqs.~(\ref{djs}) and
(\ref{hydrolin1}) we find five equations for five unknowns. Excluding
$\delta P$, $\mu_I$, and $\delta j_x$, we are left with two equations
for $u_y$ and $\delta j_{I,y}$. For further analysis it is convenient
to express them in terms of dimensionless quantities
\begin{equation}
\label{qp}
q = \frac{n_{I,0} u_y}{q_0},
\quad
p = \frac{\delta j_{I,y}}{q_0},
\quad
q_0 = \frac{\omega_B\tau_{\rm dis}E}{e\tilde R},
\end{equation}
in the matrix form
\begin{equation}
\label{eq}
\widehat L
\begin{pmatrix}
q'' \cr
p''
\end{pmatrix}
=
\widehat M
\begin{pmatrix}
q \cr
p
\end{pmatrix}
+
\begin{pmatrix}
\alpha_3 \cr
p_0
\end{pmatrix}.
\end{equation}
The matrix $\widehat{L}$ comprises squares of the
recombination-related length scales
\begin{subequations}
\begin{equation}
\widehat L =
\begin{pmatrix}
\ell^2_{RG}-\ell^2_{R1} & -\ell^2_{R1}\cr
\ell^2_{R2}  & \ell^2_{R2}
\end{pmatrix},
\end{equation}
\begin{equation}
\label{lrg}
\ell^2_{RG} = \frac{1}{2}\ell^2_{RE}
+
\frac{2\pi}{9\zeta(3)}\frac{\eta v_g^4\tau_{\rm dis}}{T^3},
\quad
\ell^2_{RE} = v_g^2\tau_{RE}\tau_{\rm dis},
\end{equation}
\begin{equation}
\ell^2_{R1} = \alpha_1\alpha_3\delta_I\frac{\tilde R_B}{2\tilde R}\ell^2_{R},
\quad
\ell^2_{R} = v_g^2\tau_{R}\tau_{\rm dis},
\end{equation}
\begin{equation}
\ell^2_{R2} = \delta_I\frac{R_0}{2\tilde R}
\frac{\ell^2_{R}}{1\!+\!\tau_{\rm dis}/(\delta_I\tau_{22})},
\end{equation}
while the remaining quantities are dimensionless
\begin{equation}
\alpha_3 = \frac{2\pi^2\ln2}{27\zeta(3)}\approx 0.42,
\quad
p_0 = \frac{\alpha_1\delta_I}
{1\!+\!\tau_{\rm dis}/(\delta_I\tau_{22})},
\end{equation}
\begin{equation}
\widehat M =
\begin{pmatrix}
C_1 & 0\cr
C_2 & 1
\end{pmatrix},
\end{equation}
\begin{equation}
C_1 =\frac{R_0\!+\!\delta R(B)}{\tilde R},
\quad
C_2 = \frac{12\alpha_1\delta_I\tilde R_B\ln^22}{\pi^2\tilde R}.
\end{equation}
\end{subequations}
The correction ${\delta R(B)\propto\omega_B^2\tau_{\rm dis}}$ is
defined in Eq.~(\ref{drb}).

Once Eqs.~(\ref{eq}) are solved, we can find the electric current
(\ref{jstr}) by substituting the solutions $q(y)$ and $p(y)$ into
Eq.~(\ref{lj}) using Eqs.~(\ref{qp}) and (\ref{ceni2}). As a result,
we find
\begin{equation}
\label{jres}
\delta j_x(y) = \frac{E}{eR_0}\left[1
+
\frac{\omega_B^2\tau_{\rm dis}}{e^2\tilde R \, T}\!
\left(
\frac{\pi\alpha_1}{2\ln2}\,p
+
\frac{6\ln2}{\pi} \,
q\right)\!\right]\!.
\end{equation}
We reiterate, that Eq.~(\ref{jres}) describes {\it viscous electronic fluid}
in neutral graphene (in contrast to the inviscid system of carriers
considered in Ref.~\cite{hydro0}).

{\it Anti-Poiseuille flow} -- Equations similar to Eq.~(\ref{eq}) have
been solved in Refs.~\cite{hydro0,mr1,mr3,mr2,cfl} focusing on the
resulting magnetoresistance. In this paper, we are interested in the
spatial profile of the quasiparticle currents. Requiring the
``hard-wall'' boundary conditions
\begin{equation}
\label{bcs}
u_y(\pm W/2)=\delta j_{I,y} (\pm W/2)=0,
\end{equation}
we find the solution to Eq.~(\ref{eq}) in the form of the catenary
curve
\begin{equation}
\label{res2}
\begin{pmatrix}
q \cr
p
\end{pmatrix}
= 
\left[\frac{\cosh(\widehat{\cal K}y)}{\cosh(\widehat{\cal K}W/2)}-1\right]
\widehat{M}^{-1}
\begin{pmatrix}
\alpha_3 \cr
p_0
\end{pmatrix},
\end{equation} 
where
\[
\widehat{\cal K}^2 = \widehat{L}^{-1}\widehat{M}.
\]
Substituting the result (\ref{res2}) into Eq.~(\ref{jres}) we find the
electric current profile. The analytical expression for $\delta
j_x(y)$ contains a $y$-independent contribution inherited from the
first term in Eq.~(\ref{jres}) and the second term in Eq.~(\ref{res2})
as well as the catenary terms describing the $y$ dependence of $q$ and
$p$ from Eq.~(\ref{res2}). Following Ref.~\cite{imh}, we normalize the
current by its average value
\begin{equation}
\label{jav}
\bar{j_x} = \frac{1}{W} \!\!\int\limits^{W/2}_{-W/2}\!\! dy \, \delta j_x(y),
\end{equation}
which can be obtained by averaging the solution
(\ref{res2}) and substituting the result into
Eq.~(\ref{jres}). Averaging of Eq.~(\ref{res2}) can be performed in
the matrix form yielding
\begin{equation}
\label{res3}
\begin{pmatrix}
\bar{q} \cr
\bar{p}
\end{pmatrix}
= 
\left[\frac{\tanh(\widehat{\cal K}W/2)}{\widehat{\cal K}W/2}-1\right]
\widehat{M}^{-1}
\begin{pmatrix}
\alpha_3 \cr
p_0
\end{pmatrix}.
\end{equation}

\begin{figure}[t]
\centerline{\includegraphics[width=0.9\columnwidth]{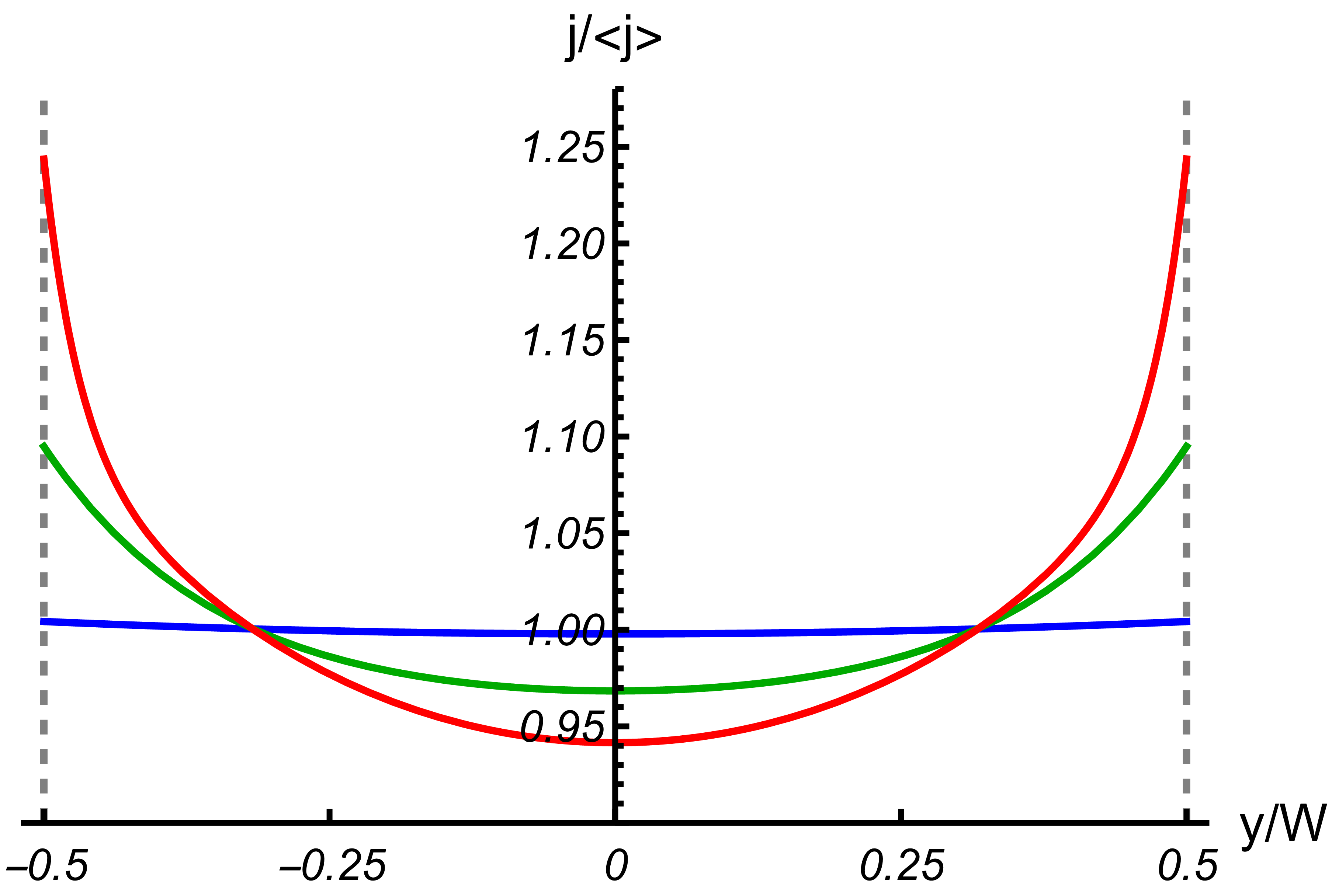}}
\caption{Catenary curves of the current density in the narrow channel
  Eq.~(\ref{jres}) normalized by the averaged current density
  Eq.~(\ref{jav}). The numerical results were obtained for typical
  parameter values (${\tau_{\rm dis}\approx0.8}\,$THz \cite{gal},
  ${\alpha_g\approx0.2}$ \cite{gal,sav}, ${\nu\approx0.4}\,$m$^2/$s
  \cite{me2,imh}, ${B=0.1}\,$T, ${T=250}\,$K) and correspond to three
  values of the channel width, ${W=0.1,\,1,\,5}\,\mu$m (blue, green,
  and red curves, respectively).}
\label{fig1:cp}
\end{figure}

The resulting inhomogeneous current density is illustrated in
Fig.~\ref{fig1:cp}. In some sense, the profile in Fig.~\ref{fig1:cp}
can be regarded as ``{\it anti-Poiseuille}'': unlike the true Poiseuille
flow, this current density exhibits a {\it minimum} in the center of
the channel and is finite at the edges (in fact, there it reaches its
maximum). The numerical values of the current density were obtained by
using a typical experimental value ${\tau_{\rm dis}\approx0.8}\,$THz
\cite{gal}, and assuming the effective coupling constant
${\alpha_g\approx0.2}$ following Refs.~\cite{gal,sav}, temperature
${T=250}\,$K, magnetic field ${B=0.1}\,$T, and channel width
${W=1}\,\mu$m. The viscosity affects the current only through the
length scale $\ell_{RG}$, see Eq.~(\ref{lrg}). This effect is rather
weak: varying the kinematic viscosity in the range
${\nu\approx0.2-0.4}\,$m$^2/$s \cite{me2} does not significantly
change the results. The recombination length ${\ell_R\approx2}\,\mu$m
and the energy relaxation length ${\ell_{RE}\approx5}\,\mu$m were
chosen phenomenologically, using the data of Ref.~\cite{meg} as a
guide (see also Ref.~\cite{meig1} for theoretical estimates).

\begin{figure}[t]
\centerline{\includegraphics[width=0.9\columnwidth]{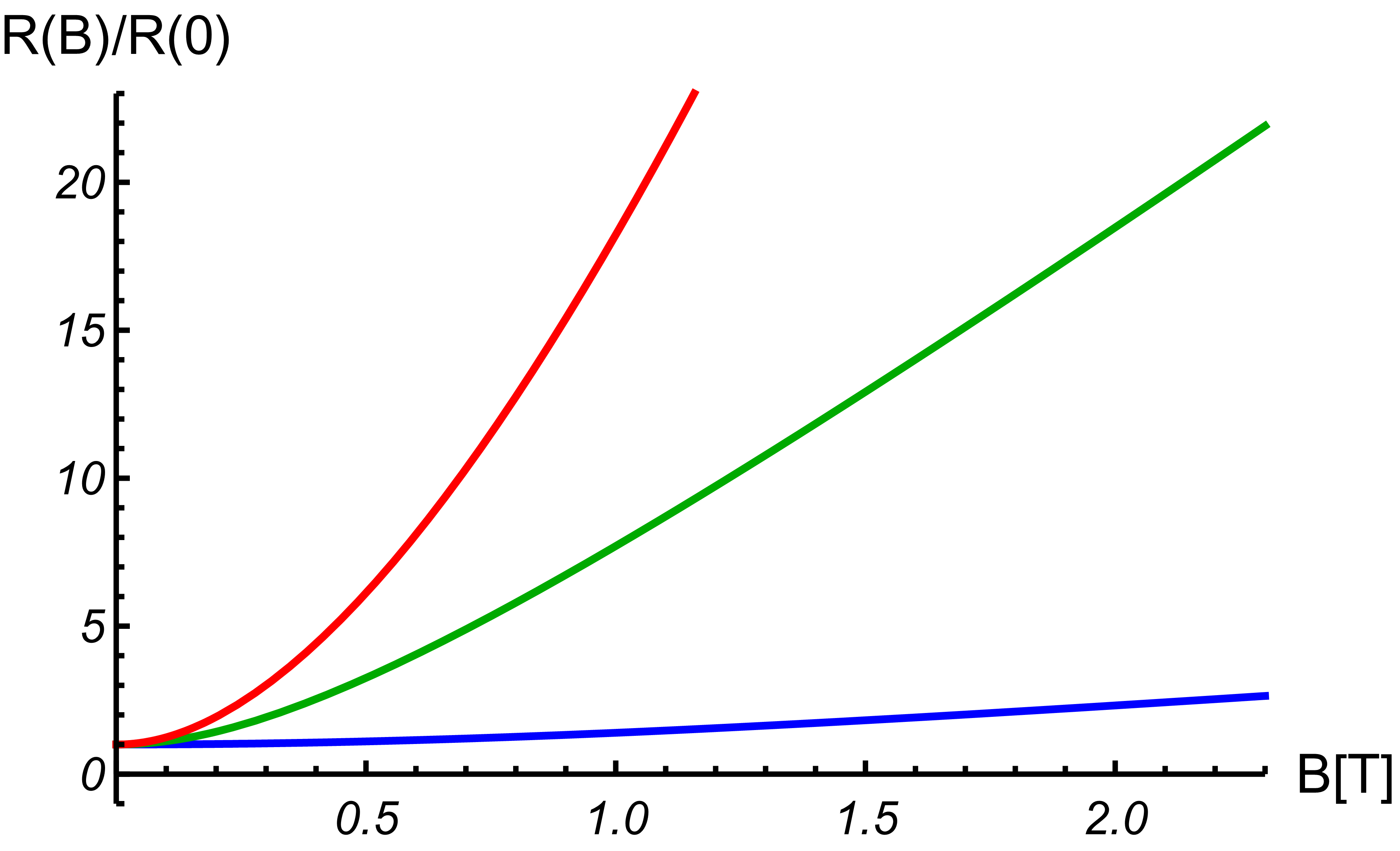}}
\caption{Magnetoresistance in the narrow channel following from
  Eqs.~(\ref{jres}) and (\ref{jav}) normalized by the zero field
  resistance $R_0$.  The numerical results were obtained for typical
  parameter values (${\tau_{\rm dis}\approx0.8}\,$THz \cite{gal},
  ${\alpha_g\approx0.2}$ \cite{gal,sav}, ${\nu\approx0.4}\,$m$^2/$s
  \cite{me2,imh}, ${B=0.1}\,$T, ${T=250}\,$K) and correspond to three
  values of the channel width, ${W=0.1,\,1,\,5}\,\mu$m (blue, green,
  and red curves, respectively).}
\label{fig2:mr}
\end{figure}

{\it Discussion} -- The results presented in this paper have to be
contrasted with recent developments in the field. Most theoretical
work on hydrodynamic behavior in neutral (or compensated) materials
has been devoted to infinite (or bulk) systems
\cite{luc,rev,kash,mus,hydro0,me1}. A bulk system is translationally
invariant and hence the current density is uniform with the
corresponding sheet resistance given by $R_0$. In confined geometries
the resulting flow profiles are determined by the interplay of sample
geometry, boundary conditions, and bulk interaction effects
\cite{kash18}. With respect to electron-electron interaction, three
types of theories have been proposed: (i) macroscopic linear response
theory of the inviscid electronic fluid \cite{hydro0}, (ii) two-fluid
hydrodynamics \cite{mr1,mr2,mr3,sven2}, and (iii) viscous electronic
hydrodynamics that is the subject of the present paper. The difference
between the three theories can be summarized as follows. (i)
Ref.~\cite{hydro0} generalized the standard transport theory
(basically the Ohm's law) to graphene close to charge neutrality,
where electron-electron interaction contributes to resistivity directly
due to lack of Galilean invariance. The resulting theory comprises
three (algebraic) equations for three macroscopic currents and does
not take into account any possible viscous effects. (ii) The two-fluid
model of Ref.~\cite{mr2} assumes that the electron and hole subsystems
(i.e. quasiparticles in two different bands) are independently
equilibrated and form two separate fluids, while the electron-hole
scattering leads to a (weak) friction between the two resembling the
drag effect \cite{dragrev}. The theory is described by two sets of
hydrodynamic equations, including two Navier-Stokes-like equations. In
contrast, (iii) the present hydrodynamic theory
\cite{rev,luc,hydro1,me1} assumes that the whole system of charge
carriers is equilibrated and is described by a single local
equilibrium distribution function leading to the generalized
Navier-Stokes equation (\ref{eq1g}).

The only theory (out of the above three) yielding the Poiseuille-like
flow for the electric current in the channel geometry in the absence
of magnetic field is the two-fluid model of Ref.~\cite{mr2}, which
assumes no-slip boundary conditions for each fluid. Neither the linear
response theory of Ref.~\cite{hydro0}, nor the theory presented in
this paper allow for this behavior. The fact that both approaches
yield qualitatively similar results (e.g., the absence of the
Poiseuille flow and linear magnetoresistance) is quite remarkable
since these are two very different theories describing two different
systems, one being a (nearly relativistic) viscous fluid and the other
being a standard, inviscid (two-band) system of charge carriers. Even
though in the latter approach viscosity as a stress-stress correlator
\cite{read,julia,bur19} might not not necessarily vanish, none of the
macroscopic currents satisfy a second-order differential equation of
the Navier-Stokes type. It is then rather natural that this approach
does not allow for a Poiseuille-like flow. In contrast, the present
theory is fully hydrodynamic and hence does in principle yield
Poiseuille-like solutions \cite{julia1}. What we have shown here is
that such flows cannot be driven by the electric field leaving the
temperature gradient \cite{julia1} as the only possibility to induce
the Poiseuille flow in neutral graphene.

All of the above references agree that in the absence of magnetic
field the electric current density is uniform not only in the bulk
(infinite) systems, but also in the channel geometry. Based on the
arguments presented in this paper, we believe that this intuitively
expected conclusion follows from implicit assumptions of either
specular boundary conditions or diffusive bulk transport (where one
typically neglects narrow regions of inhomogeneity at the sample
edges). Here we considered a narrow channel, which is no longer
translationally invariant in the lateral direction. In the special
case of specular scattering off the boundaries, we find basically the
same results: the current density (\ref{jres}) is uniform with $R_0$
being the resistance. Note, that similarly to the bulk case, $R_0$
remains finite even in the limit of a completely clean system,
$\tau_{\rm dis}\rightarrow\infty$. Once magnetic field is applied, the
bulk system exhibits \cite{hydro0,mus} positive, parabolic
magnetoresistance $\delta R(B)$, see Eq.~(\ref{drb}). In contrast, the
electronic flow constrained to the narrow channel exhibits linear
magnetoresistance \cite{hydro0} in classically strong magnetic fields,
see Fig.~\ref{fig2:mr}.

Linear magnetoresistance was also discussed in the context of the
two-fluid hydrodynamics in Refs.~\cite{mr1,mr2,mr3,sven2}. These
papers considered a phenomenological model of compensated semimetals
where elementary excitations of the conductance and valence bands,
i.e. electrons and holes, independently formed hydrodynamic flows,
which were only weakly coupled by a mutual friction term. In the
language of scattering rates, this model assumed that intraband
scattering (characterized by $\tau_{ee}$ and $\tau_{hh}$ in
self-evident notation) was much more effective that interband
scattering, such that $\tau_{eh}\gg\tau_{ee},\,\tau_{hh}$. The
zero-field resistance of this model is provided by disorder and
intraband scattering, such that even in a clean system ($\tau_{\rm
  dis}\rightarrow\infty$) the resistance is finite (and is determined
by $\tau_{eh}$ in a way that is reminiscent of Coulomb drag
\cite{dragrev,meg,df1,df2}).

We also stress the importance of boundary conditions on the
distribution function. In particular, Ref.~\cite{hydro0} considered
linear magnetoresistance in a narrow channel, but avoided the issue of
the boundary conditions altogether (moreover, energy relaxation was
considered purely phenomenologically). Based on the present results,
we conclude that the theory presented in Ref.~\cite{hydro0} is valid
for specular scattering off the channel boundaries. The two-fluid
model of Refs.~\cite{mr1,mr2,mr3,cfl} assumed hydrodynamic no-slip
boundary conditions for each of the fluids, such that the resulting
electric current would vanish at the boundaries. This approach is
justified in a different parameter regime from that of the
hydrodynamic theory of electronic transport in graphene
\cite{me1,rev,luc} with a single hydrodynamic flow. Here the electric
current comprises both the hydrodynamic and dissipative contributions
\cite{me3}, see Eq.~(\ref{jlr}). At charge neutrality, the current is
decoupled from the hydrodynamic flow and hence the hydrodynamic
boundary conditions \cite{ks19}. Instead, one should consider the
kinetics of scattering off the boundaries \cite{bee}. In the special
case of specular scattering considered in this paper, the
nonequilibrium distribution function retains the form (\ref{hs}). In
the case of diffusive scattering the distribution function is more
complicated; in both cases the boundary condition on the distribution
function does not easily translate into a boundary condition for
electric current: in particular, the electric current is not expected
to vanish at the channel boundaries \cite{imm}. The alternative
no-stress boundary condition \cite{ks19,fl0}, that could have been
chosen in the two-fluid model of Refs.~\cite{mr1,mr2,mr3,cfl}, would
not yield the results shown in Fig.~\ref{fig1:cp} as well: then the
current density profile would have been flat at the channel
boundaries.

Finally, our conclusions should be contrasted with the results of the
recent imaging experiment of Ref.~\cite{imh}. In particular, the
vanishing current density at the channel boundaries reported in
Ref.~\cite{imh} are consistent with the hydrodynamic no-slip boundary
condition that within our theory is incompatible with the charge flow
in neutral graphene. Based on the arguments presented in this paper,
as well as our preliminary results for the case of diffusive
scattering of the channel boundaries, we expect that bulk
recombination processes (most notably, supercollisions) are
responsible for the small dip in the current density seen in
Ref.~\cite{imh} in the center of the channel. The overall shape of the
current density profile reported in Ref.~\cite{imh} is consistent with
the charge flow under assumptions of the diffusive boundary conditions
(to be discussed in a subsequent publication). However, at this time
we are not aware of any theoretical argument that would predict
precise vanishing of electric current at the channel boundaries (in
particular, a recent study of hydrodynamic boundary conditions in
graphene \cite{ks19} reported a nonvanishing slip length). This point
appears even more intriguing in view of the recent experiment
demonstrating current-carrying edge states in graphene \cite{zel},
possibly a manifestation of the edge charge accumulation. The latter
physics (in particular, the role of such ``edge reconstruction'' in
the hydrodynamic regime) is yet to be addressed in a consistent
theoretical fashion. Combining the observations of Ref.~\cite{imh} and
Ref.~\cite{zel} with the peculiarities of the hydrodynamic approach
for neutral graphene remains an important open question.

To conclude, we have discussed electronic transport in graphene at
charge neutrality exhibiting a behavior that is strikingly different
from any single-component fluid including that in strongly doped
graphene. For weak doping ($\mu\ll T$), the hydrodynamic contribution
to the electric current Eq.~(\ref{jlr}) yields a small correction to
the results presented in this paper (e.g, the hydrodynamic
contribution to optical conductivity in weakly doped graphene was
shown \cite{me3} to be proportional to $\mu^2/T^2$). For $\mu\sim T$,
both the hydrodynamic and dissipative (``kinetic'') contributions are
of the same order. Now there is no small parameter in the
theory and the full system of linearized hydrodynamic equations can be
represented by a $6\times6$ matrix \cite{megt}. In the strongly doped
regime ($\mu\gg T$), the hydrodynamic contribution dominates and in
addition the boundary scattering becomes diffusive. As a result, the
electronic flow in a channel exhibits the Poiseuille profile in
agreement with the experimental observations in Ref.~\cite{imm}. Thus
we expect the crossover from the anti-Poiseuille to Poiseuille flow to
take place at $\mu\sim T$.

{\it Summary} -- In this paper we have shown that electronic flow in
neutral graphene is qualitatively different from that in a
conventional viscous fluid. Our main results can be summarized as
follows: (i) in response to external electric field, channel-shaped
samples of neutral graphene do not exhibit Poiseuille-like flows,
while the resulting electric current is independent of viscosity
regardless of the choice of the boundary conditions; (ii) for specular
boundaries, the electric current density is spatially homogeneous; but
(iii) it can be made inhomogeneous by applying the external magnetic
field. In the latter case the current profile is anti-Poiseuille, see
Fig.~\ref{fig1:cp}.

\section*{Acknowledgments} 

The authors are grateful to P. Alekseev, U. Briskot, I. Burmistrov,
A. Dmitriev, V. Gall, V. Kachorovskii, E. Kiselev, A. Mirlin,
J. Schmalian, M. Sch\"utt, and A. Shnirman for fruitful
discussions. This work was supported by the German Research Foundation
DFG within FLAG-ERA Joint Transnational Call (Project GRANSPORT), by
the European Commission under the EU Horizon 2020 MSCA-RISE-2019
Program (Project 873028 HYDROTRONICS), by the German Research
Foundation DFG project NA 1114/5-1 (BNN), by the German-Israeli
Foundation for Scientific Research and Development (GIF) Grant
No. I-1505-303.10/2019 (IVG) and by the Russian Science Foundation,
Grant No. 20-12-00147 (IVG). BNN acknowledges the support by the MEPhI
Academic Excellence Project, Contract No. 02.a03.21.0005.


\appendix*


\section{Dissipative corrections to macroscopic currents}
\label{djapp}

Within the three-mode approximation \cite{me1}, the hydrodynamic
theory in graphene is formulated in terms of three macroscopic
currents (\ref{jlr}). In local equilibrium, all three currents are
proportional to the hydrodynamic velocity $\bs{u}$. The effect of
electron-electron interaction beyond local equilibrium is captured by
the dissipative corrections that can be found following the standard
perturbative approach \cite{dau6}. In the context of electronic
hydrodynamics in graphene, the dissipative corrections were derived in
Refs.~\cite{me1,hydro1,me3}. Here we present a slightly modified
approach better suited for the problem at hand.

Let us highlight the main differences between the electronic
hydrodynamics in graphene and the conventional hydrodynamics of
Galilean-invariant fluids: (i) the band structure of graphene contains
two bands touching at the Dirac points leading to the presence of two
types of carriers characterized by two quasiparticle currents,
$\bs{j}$ and $\bs{j}_I$; (ii) neither of the two currents represent
the flow of momentum described by the energy current $\bs{j}_E$; (iii)
charge carriers in graphene may scatter off lattice imperfections
(impurities), lattice vibrations (phonons), and experience other
scattering processes leading to violation of conservation laws
including momentum conservation.

Due to the latter issue, the hydrodynamic approach to electronic
transport in graphene (as well as any other solid) may be justified
only in an intermediate temperature regime, where the
electron-electron interaction is the dominant scattering process
characterized by the largest relaxation rate or the smallest timescale
\cite{rev,luc}
\[
\tau_{ee} \ll \tau_{\rm dis}, \tau_R, \,\, {\rm etc}.
\]
Local equilibrium is formed at the shortest timescales of the order
of $\tau_{ee}$. As pointed out in Ref.~\cite{hydro0}, in graphene this
local equilibrium is not equivalent to a steady state since the
electron-electron interactions do not relax momentum and hence the
hydrodynamic energy flow. To overcome this difficulty one has to take
into account weak disorder scattering leading, e.g., to parabolic
magnetoresistance \cite{mus,hydro0}. We emphasize that disorder
scattering contributes to the hydrodynamic theory already at local
equilibrium \cite{me1}. Technically this can be understood from the
fact that the local equilibrium distribution function does not nullify
the disorder collision integral. Similarly, local equilibrium in
graphene is affected by electron-phonon scattering
\cite{hydro0,me1,meig1,alf,lev19,fos16,meg}. Since the lowest-order
electron-phonon scattering is kinematically suppressed (within the same
valley), the dominant process appears to be the disorder-assisted
electron-phonon scattering (or supercollisions) \cite{srl,meig1}. As
compared to the direct impurity scattering, these processes are
second-order. Nevertheless, we assume that the mean free time
$\tau_{\rm dis}$ includes the (small) contribution of supercollisions
as well. The more important effect of supercollisions are the weak
decay terms in the continuity equations for the energy and imbalance
densities, Eqs.~(\ref{eqen}) and (\ref{ceni1}) that are characterized
by the timescales $\tau_{RE}$ and $\tau_R$ \cite{meig1}. Again, these
effects appear already at local equilibrium.

Within linear response, the local equilibrium state we have described
so far is fully equivalent \cite{me1} to the standard transport theory
yielding the Ohm's law, classical Hall effect, and -- at charge
neutrality -- positive, parabolic magnetoresistance. As such, the
hydrodynamic theory already includes the dissipative processes related
to the weak disorder and electron-phonon coupling. This point
represents the most important difference between electronic
hydrodynamics and conventional fluids, where the ideal flow is always
isentropic \cite{dau6}. In the latter case, dissipative processes
(viscosity and thermal conductivity) are attributed to the same
interparticle collisions that are responsible for equilibration. By
analogy, the effect of electron-electron interaction in electronic
hydrodynamics beyond local equilibrium is also described in terms of
the ``dissipative corrections'' to quasiparticle currents (as well as
viscosity), the term that might cause confusion (since some
dissipation is already taken into account). Moreover,
electron-electron interaction does not lead to any further correction
to the energy current (since it conserves momentum). It is therefore
logical to consider two corrections $\delta\bs{j}$ and
$\delta\bs{j}_I$ due to electron-electron interaction instead of three
introduced in Ref.~\cite{me1}.

To describe the dissipative processes beyond local equilibrium one
introduces a nonequilibrium correction to the local equilibrium
distribution function $f_{\lambda\bs{k}}^{(0)}$ \cite{dau10}
\begin{equation}
\label{df}
\delta f_{\lambda\bs{k}} = f_{\lambda\bs{k}} \!-\! f_{\lambda\bs{k}}^{(0)}
= -T\frac{\partial f_{\lambda\bs{k}}^{(0)}}{\partial\epsilon_{\lambda\bs{k}}} h_{\lambda\bs{k}}
= f_{\lambda\bs{k}}^{(0)}\!\left(1\!-\!f_{\lambda\bs{k}}^{(0)}\right)\! h_{\lambda\bs{k}},
\end{equation}
where the single-particle states are labeled by the band index
${\lambda=\pm}$ and the momentum $\bs{k}$. Taking advantage of the
so-called collinear scattering singularity in graphene
\cite{msf,mfss,kash,mus,bri,schutt,drag2,hydro0,hydro1,me1}, we adopt
the ``three-mode approximation'' \cite{hydro0,hydro1,me1} and write
the correction $h$ in the form 
\begin{subequations}
\label{hs}
\begin{equation}
\label{hs0}
h_{\lambda\bs{k}} = \frac{\bs{v}_{\lambda\bs{k}}}{v_g}\sum_1^3 \phi_i \bs{h}^{(i)}
+ \frac{v^\alpha_{\lambda\bs{k}} v^\beta_{\lambda\bs{k}}}{v_g^2}  \sum_1^3 \phi_i h_{\alpha\beta}^{(i)} + \dots,
\end{equation}
where $\dots$ stands for higher-order tensors and the ``three modes''
are expressed by means of ($\epsilon_{\lambda\bs{k}}$ denotes the
quasiparticle spectrum)
\begin{equation}
\label{phis}
\phi_1 = 1, \quad \phi_2 = \lambda, \quad \phi_3 = \epsilon_{\lambda\bs{k}}/T.
\end{equation}
The first term in $h$ is responsible for dissipative corrections to
the currents, the second term -- for viscosity \cite{me1}.

The coefficients $\bs{h}^{(i)}$ and $h_{\alpha\beta}^{(i)}$ in
Eq.~(\ref{hs0}) satisfy general constraints \cite{dau10} reflecting
the postulate that electron-electron collisions should not alter
conserved thermodynamic quantities. To maintain conservation of the
number of particles and energy one sets \cite{hydro1,me1}
\begin{equation}
\label{trh0}
{\rm Tr}\, h_{\alpha\beta}^{(i)}=0.
\end{equation}
To maintain momentum conservation, we require that any correction to
the energy current due to the nonequilibrium correction (\ref{df})
should vanish leading to 
\begin{equation}
\label{h3}
\bs{h}^{(3)} = -\frac{2T}{3n_E}\left(n\bs{h}^{(1)}+n_I\bs{h}^{(2)}\right),
\end{equation}
\end{subequations}
following from the linear correspondence between the coefficients
$\bs{h}^{(i)}$ and the corrections to the currents
\cite{hydro1,me1}
\begin{equation}
\label{djso}
\begin{pmatrix}
\delta\bs{j} \cr
\delta\bs{j}_I \cr
\delta\bs{j}_E/T
\end{pmatrix}
=
\frac{v_gT}{2} 
\widehat{M}_h\!
\begin{pmatrix}
\bs{h}^{(1)} \cr
\bs{h}^{(2)} \cr
\bs{h}^{(3)}
\end{pmatrix},
\end{equation}
where 
\begin{equation}
\label{mh}
\widehat{M}_h
=
\begin{pmatrix}
 \frac{\partial n}{\partial\mu} & \frac{\partial n_{I}}{\partial\mu} & \frac{2n}{T} \cr
 \frac{\partial n_{I}}{\partial\mu} & \frac{\partial n}{\partial\mu} & \frac{2n_{I}}{T} \cr
 \frac{2n}{T} & \frac{2n_{I}}{T} & \frac{3n_{E}}{T^2} \cr
\end{pmatrix}.
\end{equation}
Enforcing the constraint (\ref{h3}) we find ${\delta\bs{j}_E=0}$,
while for the remaining two dissipative corrections we obtain
\begin{subequations}
\label{djsc}
\begin{equation}
\label{djc}
\delta\bs{j} \!=\! \frac{v_gT}{2} \!
\left[\!
\left(
\frac{\partial n}{\partial\mu} 
\!-\!
\frac{4n^2}{3n_E}
\right)\!
\bs{h}^{(1)} 
\!\!+\!
\left(
\frac{\partial n_{I}}{\partial\mu} 
\!-\!
\frac{4nn_I}{3n_E}
\right)\!
\bs{h}^{(2)}\!
\right]\!\!,
\end{equation}
\begin{equation}
\label{djIc}
\delta\bs{j}_I \!=\! \frac{v_gT}{2} \!
\left[\!
\left(
\frac{\partial n_I}{\partial\mu} 
\!-\!
\frac{4nn_I}{3n_E}
\right)\!
\bs{h}^{(1)} 
\!\!+\!
\left(
\frac{\partial n}{\partial\mu} 
\!-\!
\frac{4n_I^2}{3n_E}
\right)\!
\bs{h}^{(2)}\!
\right]\!\!,
\end{equation}
\end{subequations}
At charge neutrality these expressions simplify to
\begin{subequations}
\label{djsc0}
\begin{equation}
\label{djc0}
\delta\bs{j} = \frac{v_gT}{2}
\frac{\partial n}{\partial\mu} 
\bs{h}^{(1)} ,
\end{equation}
\begin{equation}
\label{djIc0}
\delta\bs{j}_I = \frac{v_gT}{2} 
\frac{\partial n}{\partial\mu} 
\delta_I
\bs{h}^{(2)},
\end{equation}
where
\begin{equation}
\label{di}
\delta_I = 1-\frac{\pi^4}{162\zeta(3)\ln2}
\approx 0.28,
\end{equation}
\end{subequations}
and $\zeta(z)$ is the Riemann's zeta function.

The approach described so far is fully justified in bulk (or infinite)
systems where one may assume rotational invariance. In contrast, if
the electronic system is confined to a narrow channel, then the
specific form of the nonequilibrium distribution function (\ref{hs})
cannot be assumed on symmetry grounds. Instead, one should solve the
kinetic equation in the presence of the boundaries imposing proper
boundary conditions on the distribution function reflecting physical
assumptions of the nature of electron scattering off the channel
boundaries \cite{bee}. In the case of specular scattering, the
distribution function satisfies
\begin{equation}
\label{bcf}
f(\pm W/2, \varphi) = f(\pm W/2, -\varphi),
\end{equation}
where $\varphi$ is the angle between the velocity
$\bs{v}_{\lambda\bs{k}}$ and the boundary (i.e., the direction along
the channel). One can easy convince oneself that the first term in
Eq.~(\ref{hs0}) satisfies this condition. Indeed, the vectors
$\bs{h}^{(1,2)}$ are linear combinations of the currents
$\delta\bs{j}$ and $\delta\bs{j}_I$, see Eqs.~(\ref{djsc}). The
electric current $\delta\bs{j}$ has only a component along the
channel, see Eq.~(\ref{jstr}), while the lateral component of the
imbalance current vanishes at the boundary, see Eq.~(\ref{bcs}).
Precisely at the boundary, the angular dependence of the first
term in Eq.~(\ref{hs0}) is therefore
\[
h\propto\cos\varphi.
\]
Similarly, the lateral component of the hydrodynamic velocity $\bs{u}$
vanishes at the boundary, see Eq.~(\ref{bcs}), such that the product
$\bs{u}\!\cdot\!\bs{k}$ has the same angular dependence (recall that
both velocity and momentum have the same direction). As a result, at
the boundary the full distribution function depends on $\cos\varphi$
only, thus satisfying Eq.~(\ref{bcf}).

The nonequilibrium correction to the distribution function can be
found using the standard iterative solution of the kinetic equation
\cite{dau10}. In the context of the three-mode approximation in
graphene, we may solve the kinetic equation directly in terms of the
dissipative corrections (\ref{djsc}) by integrating the kinetic
equation to obtain the macroscopic equations for the quasiparticle
currents. The iterative procedure is implemented by using the local
equilibrium distribution function in the left-hand side of the kinetic
equation, while retaining the nonequilibrium correction in the
right-hand side to the linear order \cite{hydro1,me1}. At charge
neutrality, the resulting equations have the form \cite{me1}
\begin{subequations}
\label{djeqs0}
\begin{equation}
\label{djeq0}
-\frac{v_g^2}{2}\frac{\partial n}{\partial\mu}e\bs{E}
+
\omega_B \bs{e}_B\!\times\!\bs{\cal K}
=
\bs{\cal I}_1,
\end{equation}
\begin{equation}
\label{djIeq0}
\frac{v_g^2}{2}\bs{\nabla}n_I
\!-\!
\frac{v_g^2n_I}{3n_E}\bs{\nabla}n_E
\!+\!
\frac{2ev_g^2n_I}{3cn_E} \delta\bs{j}\!\times\!\bs{B}
\!+\!
\omega_B \bs{e}_B\!\times\!\bs{\cal K}_I
\!=\!
\bs{\cal I}_2,
\end{equation}
\end{subequations}
where the Lorentz terms are given by \cite{me1}
\begin{subequations}
\label{ks}
\begin{equation}
\label{k}
\bs{\cal K}(\mu=0) = (T\ln2) \frac{\partial n}{\partial\mu} \bs{u}
+
\alpha_1\delta\bs{j}_I,
\end{equation}
\begin{equation}
\label{a3}
\alpha_1
\!=\!
\frac{1\!-\!\alpha_3}{\delta_I} \!\approx\!2.08,
\quad
\alpha_3 \!=\! \frac{4n_IT\ln2}{3n_E} \!=\! \frac{2\pi^2\ln2}{27\zeta(3)},
\end{equation}
\begin{equation}
\label{ki}
\bs{\cal K}_I(\mu\!=\!0) = \delta\bs{j}.
\end{equation}
\end{subequations}
The integrated collision integrals due to electron-electron
interaction $\bs{\cal I}_i$ were discussed in detail in
Refs.~\cite{me1,me3}. At charge neutrality
\begin{subequations}
\label{is}
\begin{equation}
\label{i1}
\bs{\cal I}_1(\mu\!=\!0) = 
-\left(\frac{1}{\tau_{11}}+\frac{1}{\tau_{\rm dis}}\right) \delta\bs{j},
\end{equation}
\begin{equation}
\label{i2}
\bs{\cal I}_2(\mu\!=\!0) = 
-\left(\frac{1}{\delta_I\tau_{22}}+\frac{1}{\tau_{\rm dis}}\right) \delta\bs{j}_I,
\end{equation}
where the corresponding timescales are determined only by temperature
and to the leading order have the form
\begin{equation}
\tau_{11(22)}^{-1}(\mu\!=\!0) = \frac{\alpha_g^2Tt^{-1}_{11(22)}}{4\pi\ln2}, 
\end{equation}
\begin{equation}
t^{-1}_{11}\approx 33.13,
\qquad
t^{-1}_{22}\approx 5.45,
\end{equation}
while the integrated collision integral due to impurity scattering is
characterized by the timescale [$\tau_{\rm tr}(\epsilon)$
is the transport scattering time]
\begin{equation}
\tau_{\rm dis}^{-1}=-
\int d\epsilon\frac{\partial f^{(0)}}{\partial\epsilon} \tau^{-1}_{\rm tr}(\epsilon).
\end{equation}
\end{subequations}
In this paper we choose the imbalance chemical potential as a
hydrodynamic variable using the relation (at charge neutrality
\cite{me1})
\begin{eqnarray}
\label{gmui}
&&
\frac{1}{2}\bs{\nabla}n_I
\!-\!
\frac{n_I}{3n_E}\bs{\nabla}n_E
=
\\
&&
\nonumber\\
&&
\qquad
=
\frac{1}{2}\frac{\partial n}{\partial\mu}
\left[1-\frac{4n_I^2}{3n_E} \frac{1}{\partial n/\partial\mu}\right] \bs{\nabla}\mu_I
=\frac{\delta_I}{2}\frac{\partial n}{\partial\mu}\bs{\nabla}\mu_I.
\nonumber
\end{eqnarray}
Resolving the equation for the imbalance current, we find
\begin{equation}
\label{ji1}
\delta\bs{j}_I = -\frac{
\delta_I\frac{v_g^2}{2}\frac{\partial n}{\partial\mu}\bs{\nabla}\mu_I
+
\omega_B (1\!-\!\alpha_3)\bs{e}_B\!\times\!\delta\bs{j}}
{\tau_{\rm dis}^{-1}\!+\!\delta_I^{-1}\tau_{22}^{-1}}.
\end{equation}
Substituting this expression into Eq.~(\ref{djeq0}), we find the
dissipative correction to the electric current
\begin{eqnarray}
\label{j1}
&&
\delta\bs{j} = 
\frac{1}
{e^2(R_0\!+\!\alpha_1^2\delta_I\tilde{R}_B)}
\Bigg[ e\bs{E} +
\\
&&
\nonumber\\
&&
\qquad
+
\frac{\alpha_1\delta_I\omega_B}{\tau_{\rm dis}^{-1}\!+\!\delta_I^{-1}\tau_{22}^{-1}}
\bs{e}_B\!\times\!\bs{\nabla}\mu_I
-\omega_B
\frac{2T\ln2}{v_g^2}
\bs{e}_B\!\times\!\bs{u}\Bigg],
\nonumber
\end{eqnarray}
where $R_0$ denotes the intrinsic resistivity \cite{kash,luc} at ${\bs{B}=0}$
\begin{equation}
\label{r0}
R_0 = \frac{\pi}{2\ln2}\frac{1}{e^2T}\left(\frac{1}{\tau_{11}}+\frac{1}{\tau_{\rm dis}}\right),
\end{equation}
and
\begin{eqnarray}
\label{trb}
\tilde{R}_B = \frac{\pi}{2e^2T\ln2}\frac{\omega_B^2}{\tau_{\rm dis}^{-1}\!+\!\delta_I^{-1}\tau_{22}^{-1}}.
\end{eqnarray}
Substituting this result into Eq.~(\ref{ji1}), we find the dissipative
correction to the imbalance current
\begin{eqnarray}
\label{jI1}
&&
\delta\bs{j}_I = 
-\frac{\delta_I}{\tau_{\rm dis}^{-1}\!+\!\delta_I^{-1}\tau_{22}^{-1}}
\frac{1}{e^2(R_0\!+\!\alpha_1^2\delta_I\tilde{R}_B)}\times
\\
&&
\nonumber\\
&&
\quad
\times\!
\left[ \alpha_1\omega_B\bs{e}_B\!\times\!\bs{E}
\!+\!
\frac{2T\ln2}{\pi}e^2R_0\bs{\nabla}\mu_I
\!+\!
\alpha_1\omega_B^2\frac{2T\ln2}{v_g^2}
\bs{u} \right]\!.
\nonumber
\end{eqnarray}

To recover the positive magnetoresistance \cite{mus,hydro0,me1} in
bulk graphene, we recall that in an infinite system all currents and
densities are uniform. In this case, the generalized Navier-Stokes
equation (\ref{eq1g}) reduces to
\begin{equation}
\label{ns0}
0=v_g^2\frac{e}{c} \delta\bs{j}\!\times\!\bs{B}
-
\frac{3n_E\bs{u}}{2\tau_{{\rm dis}}},
\end{equation}
which yields the hydrodynamics velocity
\begin{equation}
\bs{u} = -\omega_B\tau_{\rm dis} \frac{4T\ln2}{3n_E} \bs{e}_B\!\times\!\delta\bs{j}.
\end{equation}
Substituting this expression into Eq.~(\ref{j1}), we find
\begin{equation}
\label{dju}
\delta\bs{j} = \frac{\bs{E}}{eR_0+e\delta R(B)},
\end{equation}
where
\begin{eqnarray}
\label{drb}
&&
\delta R(B) 
=
\alpha_1^2\delta_I\tilde{R}_B
+
\frac{8\ln^32}{9\zeta(3)}\frac{\pi}{2e^2T\ln2}\omega_B^2\tau_{\rm dis}
\nonumber\\
&&
\nonumber\\
&&
\qquad\qquad
= \frac{\omega_B^2\tau_{\rm dis}}{2e^2T\ln2}\frac{\pi}{9\zeta(3)}
\!\left[
1\!+\!\frac{9\zeta(3)}{\pi}\frac{\alpha_1^2\delta_I}{\tau_{\rm dis}^{-1}\!+\!\delta_I^{-1}\tau_{22}^{-1}}
\right]
\nonumber\\
&&
\nonumber\\
&&
\qquad\qquad
=
{\cal C} \frac{v_g^4B^2\tau_{\rm dis}}{c^2T^3} ,
\end{eqnarray}
with
\[
{\cal C} \approx 
\frac{1.71\!+\!1.04 \frac{\tau_{\rm dis}}{\tau_{22}}}{1\!+\!3.59\frac{\tau_{\rm dis}}{\tau_{22}}}
\underset{\tau_{\rm dis}\rightarrow\infty}{\longrightarrow}
\frac{\pi}{9\zeta(3)}\approx0.29.
\]
The positive, parabolic magnetoresistance (\ref{drb}) in bulk graphene
was previously found in this form in Refs.~\cite{hydro0,me1} and in
Ref.~\cite{mus} (where the limiting value of ${\cal C}$ was first
obtained in the two-mode limit, $\tau_{\rm
  dis}/\tau_{22}\rightarrow\infty$).

\bibliography{viscosity_refs}

\end{document}